\documentclass[final]{svjour3}
\usepackage{graphicx}
\usepackage{rotating}
\usepackage{amssymb}
\usepackage{mathptmx}
\usepackage[numbers]{natbib}
\usepackage{amsmath}
\makeatletter
\journalname{Journal of Low Temperature Physics}

\bibpunct{}{}{,}{s}{}{,}

\begin{document}

\newcommand{\hdblarrow}{H\makebox[0.9ex][l]{$\downdownarrows$}-}
\title{Simultaneous Noise and Impedance Fitting to Transition-Edge Sensor Data using Differential Evolution}

\author{A. P. Helenius \and T. A. Puurtinen \and K. M. Kinnunen \and I. J. Maasilta}

\institute{Nanoscience Center, Department of Physics, University of Jyvaskyla,\\ Jyvaskyla, FI-40014, Finland\\ Tel.: +358-40-835-3471\\ Fax:\\
\email{ari.p.helenius@jyu.fi}}

\maketitle

\begin{abstract}
We discuss a robust method to simultaneously fit a complex model both to the complex impedance and the noise data for transition-edge sensors (TES). It is based on a differential evolution (DE) algorithm, providing accurate and repeatable results with only a small increase in computational cost compared to the standard least squares (LS) fitting method. Test fits are made using both DE and LS methods, and the results compared with previously determined best fits, with varying initial value deviations and limit ranges for the parameters. 
The robustness of DE is demonstrated with successful fits even when parameter limits up to a factor of 5 from the known values were used. It is shown that the least squares fitting becomes unreliable beyond a 10\% deviation from the known values.


\end{abstract}

\section{Introduction}
Transition-edge sensor (TES) is a versatile, state-of-the-art radiation detector \cite{enss,ullomnew}, currently used in many applications, such as Particle Induced X-ray Emission spectroscopy \cite{pixe,pixe2},  and ground- and space-borne telescopes \cite{inoue,doriese}. 
However, the modelling of transition-edge sensors and finding fits to data has sometimes proven quite challenging in practice, due to the complexities of the thermal circuit of the device \cite{kinnunen,mikkomodel,lindeman,goldie,taralli,nasamodel}, as two and three block thermal models\cite{maasilta} need to be employed at times. 
Fitting these models by the standard least squares fitting or with certain initial guesses manually, as was done in references \cite{kinnunen,mikkomodel}, has proven to be tedious or even unreliable. 

Here, we propose a different approach to fit TES thermal models, which is independent of the initial parameters given, and can fit both the complex impedance and the noise data simultaneously, even for three-block models,
producing more reliable results than the least squares method. 
It is based on the differential evolution (DE) algorithm \cite{storn}, a branch of genetic algorithms. In this study, we fit previously measured data from Ref. \cite{kinnunen} both with the DE algorithm and with standard least squares \cite{lst},  and use the published manual fits in that paper as the control to evaluate the performance.  

\section{Three-block thermal model}
In Ref. \cite{kinnunen}, it was shown that for good fitting of the complex impedance and noise data,  three-block thermal models had to be employed. The model chosen for study here is the so called IH model of Ref. \cite{maasilta}, see Fig. 1. In the IH model, in addition to the heat capacity of the TES sensor element, $C_{tes}$, there are two additional heat capacities, one intermediate, $C_{2}$, and one hanging, $C_{1}$. The full equations for the complex impedance and for all the noise terms are lengthy, and can be found in full detail from Ref. \cite{maasilta}.

\begin{figure}[!tb]
\centering
\includegraphics[width=0.3\textwidth]{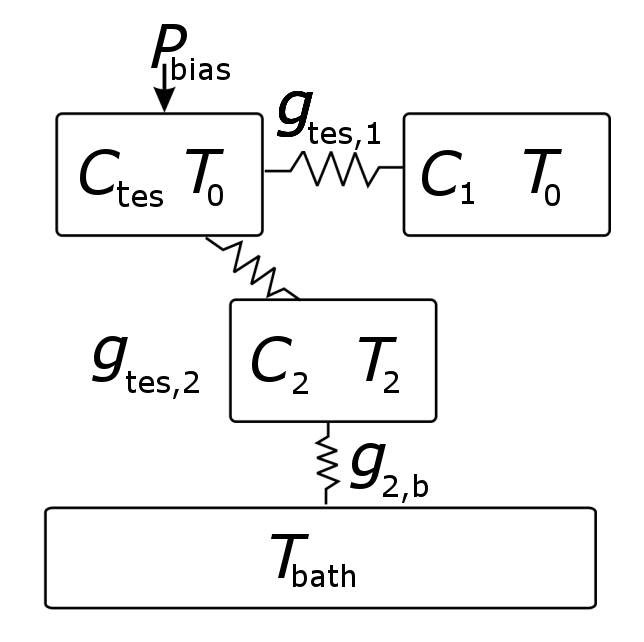}
\caption{The three-block thermal model used in this study, with additional intermediate, $C_{2}$, and hanging, $C_{1}$, heat capacities. Parameters $g_i$ describe the differential thermal conductances connecting the heat capacities, and $T_i$ are the steady state temperatures of the blocks. \label{threeblock}}
\end{figure}

\section{Genetic algorithm and differential evolution}
Differential Evolution (DE) is a high performance, yet simple, optimizer algorithm based on mutation and crossover of the trial argument vectors of the optimized function.\cite{storn} DE is initialized by selecting a number of initial population entities (typically $D$-dimensional vectors $x_i$) that are evaluated with the cost function. In each step, trial vectors are randomly mutated $y_i = x_i + F(x_j - x_k)$, where $F\in[0,2]$ is the mutation factor, and a crossover is performed by randomly mixing the vector elements $(y_i)_k \rightarrow (y_j)_k$ of two distinct entities $y_i$ and $y_j$. The cost function is then re-evaluated for the decision whether the trial is kept or discarded from the population.

Three different functions are simultaneously fitted to the measured data. The complex impedance is broken down to its real and imaginary parts and used as the first two fit functions. The third function comes from the total TES current noise, including a constant (4\,pA/$\sqrt{\text{Hz}})^2$ SQUID noise component. The cost function to be minimized for both the DE and LS fitting methods is the sum of squares of the errors of absolute values $|$fit value - data$|$, summed over the real and imaginary parts of the complex impedance and the noise. For the noise data and fit, and additional log$_{10}$ is taken 
prior to the subtraction. All the calculations are done with Python 3.7, NumPy version 1.15.1, SciPy version 1.1.0 and both of the optimization algorithms are from the package scipy.optimize: \emph{least\_squares} and \emph{differential\_evolution}. The least squares fits are done with the trust region reflecting method \cite{yuan}. The DE strategy was "best2bin", and following parameters were used for the algorithm: population size 15, mutation 1.8, recombination 0.1, tolerance $10^{-7}$, and absolute tolerance 0. No seed was was chosen, in order to see if the fit found is always the same. Polishing was used, which runs \emph{scipy.optimize.minimize} with limited memory Broyden-Fletcher-Goldfarb-Shanno algorithm and initial population was determined by "latinhypercube".

\section{Calculations and results}

A total of six fitting parameters were chosen to be free parameters for the fitting tests: All the three heat capacities of the model, the two steady state temperatures $T_0$ and $T_2$, and one of the thermal conductances, $g_{tes,2}$. $g_{tes,1}$ was kept fixed at the values determined in Ref. \cite{kinnunen}, and $g_{2,b}$ is not free anymore, if  $g_{tes,2}$, $T_0$ and $T_2$ are set,  as the overall dynamic conductance to the bath is known from the I-V measurements \cite{maasilta}.

\begin{figure}[!htb]
\centering
\includegraphics[width=\textwidth]{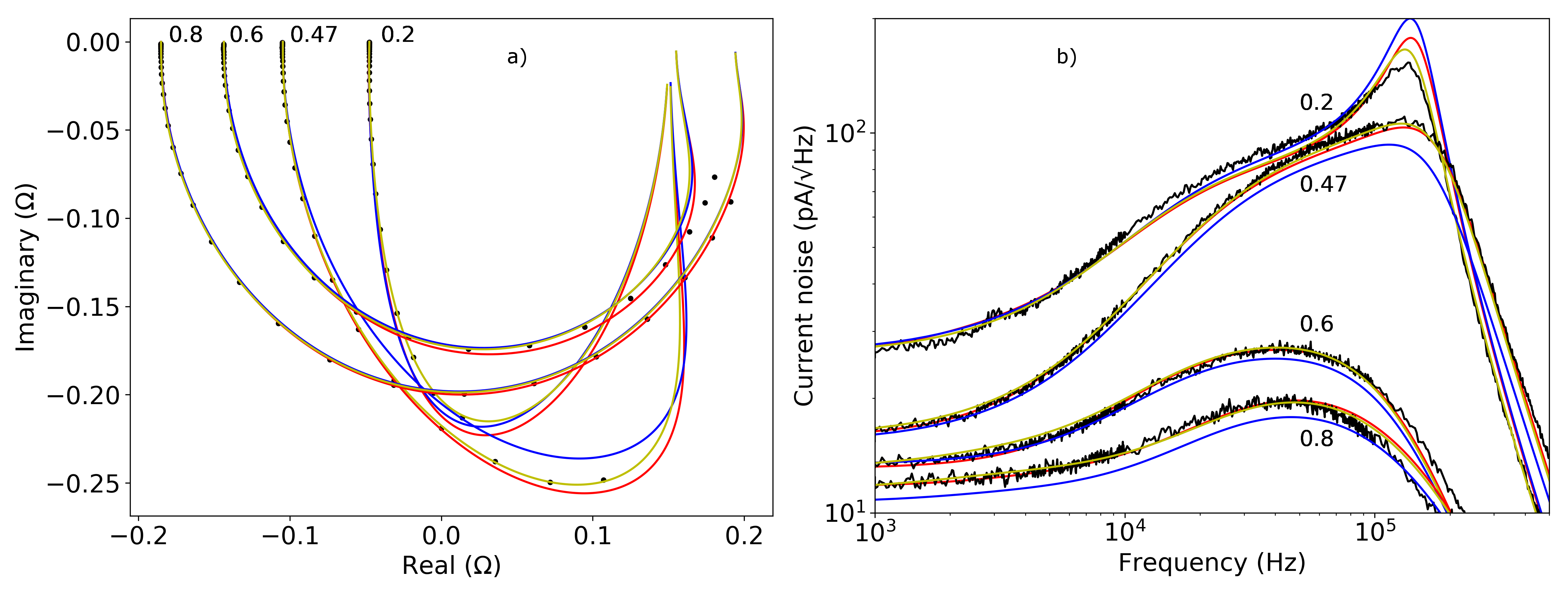}
\caption{(color online): (a) Impedance fits (curves) and data (points). Data is only taken up to 100\,kHz, fits shown up to 1\,MHz.   (b) Noise fits (colored curves) and data (black). The electrical cut off due to the measurement circuitry is visible above 100\,kHz frequencies. Manual fits (yellow lines), DE (red lines), and LS (blue lines), with 10\% deviation in limits. The bias points range from 0.2 to 0.8 $R/R_N$. \label{noisez}}
\end{figure}

For the estimation of initial values, the Corbino geometry of the devices of Ref. \cite{kinnunen} allows for a reasonable estimation of the heat capacities $C_{tes}$ and $C_1$, 
but very little is known beforehand on $C_2$, the "excess" heat capacity. The TES ($T_{0}$) and the intermediate block ($T_{2}$) temperatures have certain limits that they follow, but are typically not exactly known. A good initial guess for the TES temperature can be calculated from the I-V curves, based on the bias point dissipated power $P = IV$ and its measured bath temperature $T_{bath}$ dependence, $P = K(T_0^n-T_{\text{bath}}^n)$,  by 
$T_\text{0} = \big(\frac{P}{K}+T_{\text{bath}}^n\big)^\frac{1}{n}$.
The intermediate block temperature is somewhere between the TES temperature and the bath temperature, depending on the values of $g_{tes,2}$ and $g_{2,b}$.  
Fig. ~\ref{noisez} shows the obtained DE (red) and LS (blue) fits of complex impedance and noise at several bias points, with a 10 \% deviation of the parameter limits from the manual control fit \cite{kinnunen} (yellow). We see that both methods, in this case, give reasonable fits, with the DE method giving slightly better results. 

\begin{figure}[!htb]
\centering
\includegraphics[width=\textwidth]{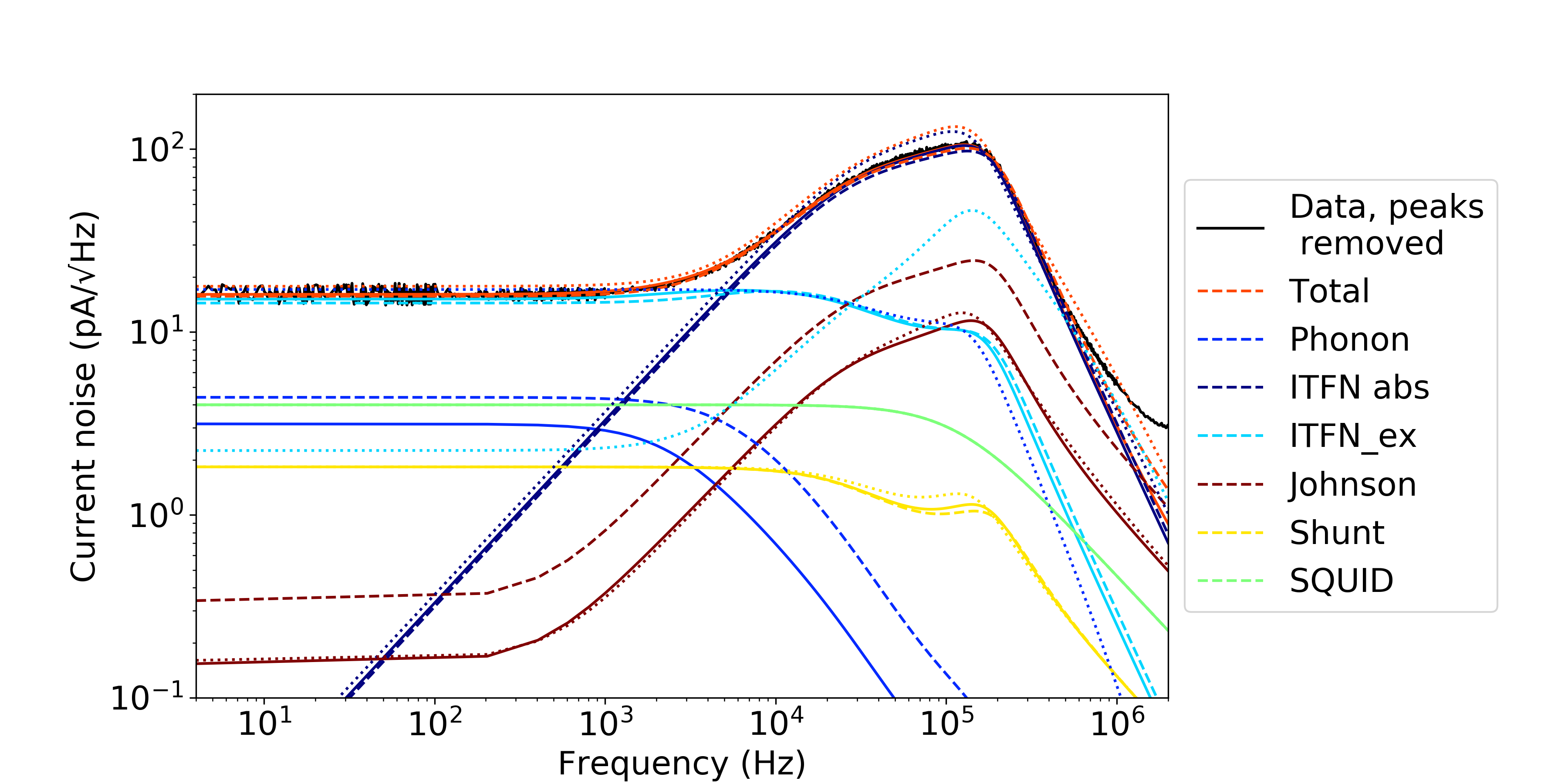}
\caption{(color online): Manual fits (solid line), DE with limits changed by a factor of five in both directions (0.2x, 5x, dashed line) and DE with limits changed by a factor of 100 in both directions (0.01x,100x, dotted line). Legend described the different noise components.    \label{noisedevs}}
\end{figure}

In Fig. ~\ref{noisedevs}, we show the effect of increasing the parameter limits significantly for the DE algorithm, in the case of noise data, to demonstrate its robustness. For the factor of five deviated DE fits, the results differ only from the manual control fit by slight changes in the Johnson and phonon noise components. Looking at each one individually, they both look as reasonable solutions, and both could be considered physical solutions. For the DE fits with factor of 100 deviated limits, the total noise fit is still reasonable, but the actual fitting values show that the TES temperature is pushed lower than the intermediate block temperature, hence rendering the solution unphysical. Thus, one should help the algorithm with all the intuition available. In this case, we could have limited the TES temperature to a small deviation from a calculated value, or in general, we could limit the fit parameters to below $\sim $ one order of magnitude from the known or estimated parameters. In this case, DE reproduces the "known" fit values to a very small margin (less than a percent), or gives another physically possible solution, as shown in Fig. ~\ref{noisedevs}.

\begin{figure}[!tb]
\centering
\includegraphics[width=\textwidth]{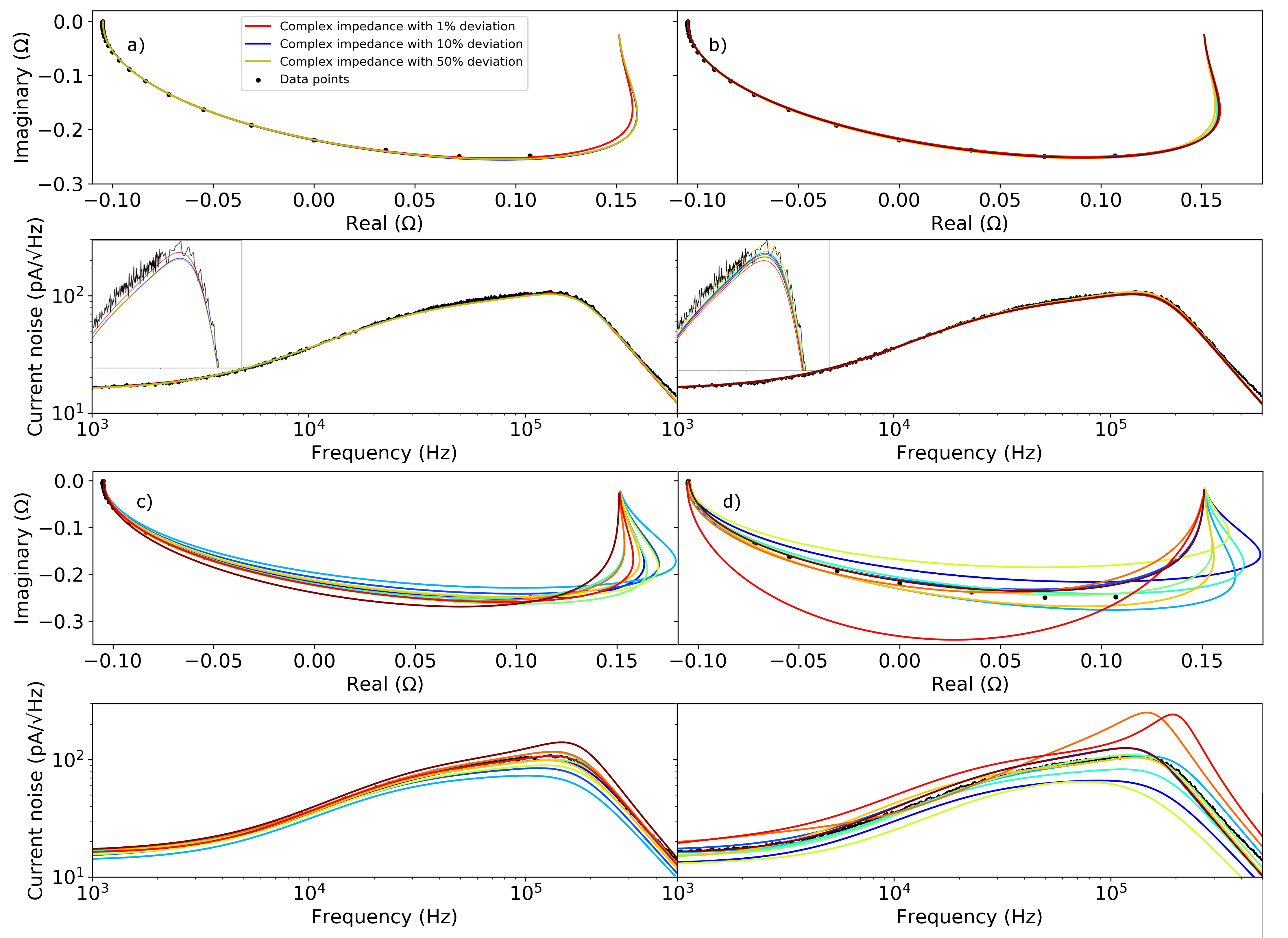}
\caption{ (color online): a) Ten randomized DE fits with 1\%,10\% and 50\% limits,  b) ten randomized LS fits with 1\%, c) 10\%, and d) 50\%  deviation from the initial value, for the same data set. Black dots and lines represent the data for impedance and noise, respectively. The upper limit for the number of iterations for both LS and DE were 5000. For DE, the number of iterations for 1\% and 10\% limits were usually around  450 and 1400, respectively, for higher deviations the maximum limit was hit. \label{multirun}}
\end{figure}

Finally, we compare both algorithms with the same lower and upper parameter limits in Fig. ~\ref{multirun}, with several different runs (initial conditions). For the least squares fit, the initial guess is randomized from the control fit parameters \cite{kinnunen}  by an additional factor of $\pm1\%$,$\pm10\%$, or $\pm20\%$ for the $1\%$,$10\%$, and $50\%$ deviated limits, respectively. As shown in Fig. \ref{multirun}, The DE algorithm is robust, but the least squares fitting does not provide reliable results for the 10 \% and 50 \% cases even with these "known" fitting parameters, as it gets easily stuck in a local minimum. This underlines the importance of the accuracy of the initial parameters for least squares, and conversely the robustness of the differential evolution algorithm. 

\section{Conclusions and Outlook}

We have implemented a robust method for simultaneous fitting of complex impedance and noise data of TES detectors by a differential evolution (DE) algorithm. The three-block thermal model has already been shown to fit this TES data well manually, but with DE, the problem of choosing the initial parameters and the tediousness and unreliability of the fitting process is removed. When the number of fitting parameters is small and the limits are close to the initial or actual physical values, least squares fitting will be faster, finding the solution in a matter of seconds, whereas DE takes roughly a minute. However, a few extra minutes, in the case of less accurate initial estimates, is a small price to pay for the robustness and reliability that DE offers. In addition, when running the calculations with larger deviations, multiple different solutions may arise as was shown in Fig. ~\ref{noisedevs}. This may require user intervention, in some cases, to drive the system towards a more physical solution. Nevertheless, DE based fitting algorithms can help avoid some of the caveats commonly encountered in multivariable non-linear fitting problems, and in particular, it is a reasonable tool for the data analysis of TES detectors.
     
%


\end{document}